\documentclass{iaus}
\usepackage{graphicx}

\title[Early-type barred galaxies] 
{Bulge properties and dark matter content of early-type barred galaxies}

\author[Corsini]  
{E.~M. Corsini} 

\affiliation{Dipartimento di Astronomia, Universit\`a di Padova, 
vicolo dell'Osservatorio 3, I-35122 Padova, Italy 
\break email: enricomaria.corsini@unipd.it}

\pubyear{2007}
\volume{245}  
\pagerange{119--126}
\date{?? and in revised form ??}
\jname{Formation and Evolution of Galaxy Bulges}
\editors{M. Bureau, E. Athanassoula \& B. Barbuy, eds.}

\begin{document}

\def\om{\Omega_B}
\def\len{a_B}
\def\lag{D_L}
\def\vpd{{\cal R}}
\def\kin{{\cal V}}
\def\pin{{\cal X}}

\maketitle

\begin{abstract}

The dynamics of a barred galaxy depends on the pattern speed of its
bar. The only direct method for measuring the pattern speed of a bar
is the Tremaine-Weinberg technique.
This method is best suited to the analysis of the distribution and
dynamics of the stellar component. Therefore it has been mostly used
for early-type barred galaxies. Most of them host a classical bulge.
On the other hand, a variety of indirect methods, which are based on
the analysis of the distribution and dynamics of the gaseous
component, has been used to measure the bar pattern speed in late-type
barred galaxies.
Nearly all the measured bars are as rapidly rotating as they can
be. By comparing this result with high-resolution numerical
simulations of bars in dark matter halos, it is possible to conclude
that these bars reside in maximal disks.

\keywords{galaxies: kinematics and dynamics, galaxies: structure}
\end{abstract}

\firstsection 

\section{Introduction}

More than $60\%$ of all bright nearby galaxies are disk-dominated
systems. Bars are seen in about $60\%$ of them. This is also true at
higher redshift. In fact, the bar fraction is constant out to $z\sim1$
(see \cite[Marinova \& Jogee 2006]{Marinova2006} and references therein).
Therefore bars are a common feature in the central regions of disk
galaxies. Their growth is partly regulated by the exchange of angular
momentum with the stellar disk and the dark matter (DM) halo. For this
reason the dynamical evolution of bars can be used to constrain the
content and distribution of DM in the inner regions of galaxy disks.

The morphology and dynamics of a barred galaxy depend on the pattern
speed of the bar, $\om$. Usually, it is parametrized with the bar
rotation rate $\vpd\equiv\lag/\len$. This is the distance-independent
ratio between the corotation radius, $\lag = V_{\rm c}/\om$ (where
$V_{\rm c}$ is the disk circular velocity), and the length of bar
semi-major axis, $\len$.
At $\lag$ the gravitational and centrifugal forces cancel out in the
rest frame of the bar.
As far as the value of $\vpd$ concerns, if $\vpd < 1.0$ the stellar
orbits are elongated perpendicular to the bar and the bar dissolves.
For this reason, self-consistent bars cannot exist in this
regime. Bars with $\vpd \gtrsim 1.0$ are close to rotate as fast they
can, and there is not a priori reason for $\vpd$ to be significantly
larger than 1.0. The knowledge of $\vpd$ allows to distinguish between
fast bars ($1.0 \leq \vpd \leq 1.4$) and slow bars ($\vpd >
1.4$). This choice does not imply anything about the actual rotation
velocity of bar.
 
\firstsection

\section{Measuring the bar pattern speed}

A variety of indirect methods has been used to measure $\om$ and
corresponding $\vpd$. They are based on the identification of
morphological features with the location of Lindblad's resonances,
comparison of the observed gas velocity and density fields with
numerical models of gas flows, and analysis of the offset and shape of
dust lanes which traces the location shocks in the gas flows. All
these methods are model dependent and are mostly used for late-type
barred galaxies because they rely on the presence of gas. Nearly all
the measured bars are found to be fast (see
\cite[Elmegreen 1996]{Elmegreen1996} for a review).

A model-independent method for measuring $\om$ is the
Tremaine-Weinberg method (\cite[Tremaine \& Weinberg
1984]{Tremaine1984}, TW).
They showed that if a tracer satisfies the continuity equation, then
$\om$ is determined from two observationally accessible quantities:
the luminosity-weighted mean of positions $\pin$ and the
luminosity-weighted mean of line-of-sight velocities $\kin$ of the
tracer. It is
\begin{equation}
\pin\,\om\,\sin i = \kin 
\end{equation}
with 
\begin{equation}
\pin = \frac{\int^{+\infty}_{-\infty}\,X\,\Sigma(X,Y)\,dX}
  {\int^{+\infty}_{-\infty}\,\Sigma(X,Y)\,dX} 
\qquad\ {\rm and} \qquad   
\kin = \frac{\int^{+\infty}_{-\infty}\,V_{\rm los}(X,Y)\,\Sigma(X,Y)\,dX}
  {\int^{+\infty}_{-\infty}\,\Sigma(X,Y)\,dX}.
\nonumber
\end{equation}
where $X$ and $Y$ are the Cartesian coordinates chosen to be aligned
with the observed major and minor axis of the galaxy disk. The
position $X$, velocity $V_{\rm los}$, and surface brightness $\Sigma$
of the tracer are measured along a strip which is parallel to the disk
major axis and offset by $Y$. This strip actually corresponds to the
aperture of either a slit or a pseudo-slit in long-slit and
integral-field spectroscopy, respectively.
Plotting $\kin$ versus $\pin$ for the different strips produces a
straight line with slope $\om \sin i$. The inclination $i$ of the
galaxy disk is derived from the analysis of the surface brightness
distribution.

The assumption underlying the TW method is that the observed surface
brightness is proportional to the surface density of the tracer, as
for old stellar populations in the absence of significant and patchy
obscuration of dust. Thus, the method has been applied largely to
absorption-line spectra of early-type barred galaxies (Table 1).
All the objects listed in Table~1 are bright galaxies, except for
NGC~4431. \cite{Corsini2007} found that the probability that the bar
of this dwarf galaxy ends close to its corotation radius (i.e., it is
fast) is about twice as likely as that the bar is much shorter than
corotation radius (i.e., it is slow).
A fast bar in NGC~4431 would suggest a common formation mechanism of
the bar both in bright and dwarf galaxies. If their disks were
previously stabilized by massive DM halos, bars were not produced by
tidal interactions because they would be slowly rotating
(\cite[Noguchi 1999]{Noguchi1999}).
But, this is not the case even for galaxies which show signs of weak
tidal interaction with a close companion (e.g., NGC~4431 and
NGC~1023). This finding implies there is no difference between TW
measurements of the stellar component in isolated or mildly
interacting barred galaxies. All the measured bars are consistent with
being in fast. This is particularly true when galaxies with small
uncertainties on $\vpd$ (e.g., $\Delta \vpd / \vpd<0.3$) are
considered.

Most of the studied SB0 bulges have the photometric and kinematic
properties of classical bulges (\cite[Aguerri et
al. 2005]{Aguerri2005}). In particular, they follow the Faber-Jackson
correlation, lie on the fundamental plane and those for which stellar
kinematics are available rotate as fast as the bulges of unbarred
galaxies. The remaining bulges have disk properties typical of
pseudobulges.

A successful application of the TW method to the stellar component of
late-type barred galaxies will remedy the selection bias present in
the current sample of directly measured pattern speeds. This problem
could be addressed by using near-infrared spectroscopy in order to
deal with dust obscuration. In fact, according to \cite{Gerssen2007}
realistic values of dust attenuation change the observed $\om$
by less than 30 per cent. 

Extensions of the TW method to multiple pattern speeds were proposed
by \cite{Corsini2003}, \cite{Maciejewski2006}, and
\cite{Merrifield2006}. They can be applied to measure pattern speeds of nested
bars, as in NGC~2950. The primary bar of this double-barred galaxy has
a smaller pattern speed than the secondary one, which is possibly
counterrotating (\cite[Corsini et al. 2003]{Corsini2003};
\cite[Maciejewski 2006]{Maciejewski2006}). 

In most cases the presence of shocks, conversion of gas between
different phases, and star formation on short timescales prevent the
application of the TW method to gas. These limitations were explored
by \cite{Rand2004} and \cite{Hernandez2005} with numerical
experiments. Ionized (Hernandez et 2004, 2005; \cite[Emsellem et
al. 2006]{Emsellem2006}; \cite[Fathi et al. 2007]{Fathi2007}), atomic
(\cite[Bureau et al. 1999]{Bureau1999}), and molecular gas
(\cite[Zimmer et al. 2004]{Zimmer2004}; \cite[Rand \& Wallin
2004]{Rand2004}) were used to derive $\om$ in a number of barred
galaxies. 

\begin{table}   
\begin{center}   
\caption{Barred galaxies with $\om$ measured by applying TW method 
  to the stellar component}
\begin{small}
\begin{tabular}{llrrrrrl}   
\noalign{\smallskip}  
\hline 
\noalign{\smallskip}  
\multicolumn{1}{c}{Galaxy}  & 
\multicolumn{1}{c}{Morp. Type} &  
\multicolumn{1}{c}{$D$} &  
\multicolumn{1}{c}{$a_B$}  & 
\multicolumn{1}{c}{$\Omega_B$} & 
\multicolumn{1}{c}{$D_L$} &  
\multicolumn{1}{c}{$\cal{R}$} &
\multicolumn{1}{c}{Ref.} \\
\multicolumn{1}{c}{} & 
\multicolumn{1}{c}{} &
\multicolumn{1}{c}{(Mpc)} &    
\multicolumn{1}{c}{(arcsec)}  & 
\multicolumn{1}{c}{(km s$^{-1}$ arcsec$^{-1}$)} & 
\multicolumn{1}{c}{(arcsec)} &
\multicolumn{1}{c}{} &  
\multicolumn{1}{c}{} \\
\multicolumn{1}{c}{(1)} & 
\multicolumn{1}{c}{(2)} &
\multicolumn{1}{c}{(3)} &    
\multicolumn{1}{c}{(4)} & 
\multicolumn{1}{c}{(5)} & 
\multicolumn{1}{c}{(6)} &
\multicolumn{1}{c}{(7)} &  
\multicolumn{1}{c}{(8)} \\
\noalign{\smallskip}  
\hline 
\noalign{\smallskip}  
ESO 139-G09 & (R)SAB0$^0$(rs) & 71.9 & $17^{+6}_{-3}$ & $21.4\pm 5.8$ & $15^{+5}_{-3}$   & $0.8^{+0.3}_{-0.2}$ & A$+$03\\ 
ESO 281-G31 & SB0$^+$(rs)     & 45.2 & $11\pm1$       & $10.5\pm 4.1$ & $20^{+12}_{-4}$  & $1.8^{+1.1}_{-0.4}$ & G$+$03\\ 
IC 874      & SB0$^0$(rs)     & 34.7 & $20\pm5$       & $ 7.0\pm 2.4$ & $27^{+13}_{-7}$  & $1.4^{+0.7}_{-0.4}$ & A$+$03\\ 
NGC 271     & (R')SBab(rs)    & 50.3 & $29\pm1$       & $ 7.8\pm 4.3$ & $44^{+30}_{-16}$ & $1.5^{+1.0}_{-0.5}$ & G$+$03\\ 
NGC 936     & SB0$^+$(rs)     & 14.9 & $50$           & $ 4.8\pm 1.1$ & $69\pm15$        & $1.4^{+0.5}_{-0.4}$ & MK95\\ 
NGC 1023    & SB0$^-$(rs)     &  5.8 & $69\pm5$       & $ 5.1\pm 1.8$ & $53^{+29}_{-14}$ & $0.8^{+0.5}_{-0.3}$ & D$+$02\\ 
NGC 1308    & SB0/a(r)        & 82.4 & $12^{+2}_{-3}$ & $39.7\pm13.9$ & $9^{+5}_{-2}$    & $0.8^{+0.4}_{-0.2}$ & A$+$03\\ 
NGC 1358    & SAB0/a(r)       & 51.6 & $19\pm3$       & $ 9.3\pm 4.5$ & $23^{+19}_{-7}$  & $1.2^{+1.0}_{-0.4}$ & G$+$03\\ 
NGC 1440    & (R')SB0$^0$(rs):& 18.4 & $24^{+6}_{-5}$ & $ 7.4\pm 1.7$ & $38^{+11}_{-7}$  & $1.6^{+0.5}_{-0.3}$ & A$+$03\\ 
NGC 2503    & SBbc(r)         & 46.0 & $34$           & $ 6.6\pm 1.6$ & $45$             & $1.4\pm0.3$         & T$+$07\\ 
NGC 2950    & (R)SB0$^0$(r)   & 19.7 & $34\pm2$       & $11.2\pm 2.4$ & $32^{+9}_{-6}$   & $1.0^{+0.3}_{-0.2}$ & C$+$03\\ 
NGC 3412    & SB0$^0$(s)      & 16.0 & $31\pm3$       & $ 4.4\pm 1.2$ & $47^{+17}_{-10}$ & $1.5^{+0.6}_{-0.3}$ & A$+$03\\
NGC 3992    & SBbc(rs)        & 16.4 & $57\pm12$      & $ 5.7\pm 0.4$ & $45\pm3$         & $0.8\pm0.2$         & G$+$03\\ 
NGC 4245    & SB0/a(r):       & 15.6 & $38$           & $ 4.7\pm1.9$  & $43$             & $1.1\pm0.5$         & T$+$07\\ 
NGC 4431    & dSB0/a          & 15.0 & $22\pm2$       & $ 7.4\pm 1.8$ & $13^{+4}_{-3}$   & $0.6^{+0.2}_{-0.1}$ & C$+$07\\ 
NGC 4596    & SB0$^+$(r)      & 29.3 & $53$           & $ 3.9\pm 1.0$ & $60^{+20}_{-12}$ & $1.2^{+0.4}_{-0.2}$ & G$+$99\\
NGC 7079    & SB0$^0$(s)      & 34.0 & $25\pm4$       & $ 8.4\pm 0.2$ & $31\pm1$         & $1.2^{+0.3}_{-0.2}$ & DW04\\
\noalign{\smallskip}  
\hline
\noalign{\smallskip}  
\noalign{\smallskip}  
\noalign{\smallskip}  
\end{tabular}
\begin{minipage}{13.4cm}  
NOTE -- 
Col.(2): Morphological classification from RC3, 
         except for ESO 281-G31 (NED) and NGC~4431 (\cite[Barazza et
         al. 2002]{Barazza2002}). NGC 2950 is a double-barred 
         galaxy and the listed values refer to its primary bar. 
Col.(3): Distance obtained as $V_{{\rm CBR}}/H_0$ with $V_{{\rm CBR}}$ 
         from RC3 and $H_0=75$ km s$^{-1}$ Mpc$^{-1}$.
Col.(4): Bar length.
Col.(5): Bar pattern speed.
Col.(6): Bar corotation radius.
Col.(7): Bar rotation rate.
Col.(8): Reference papers.
\end{minipage}   
\end{small}  
\end{center}   
\end{table}

\firstsection

\section{Dark matter distribution in barred galaxies} 

Using perturbation theory \cite{Weinberg1985} predicted that a bar
would lose angular momentum to a massive DM halo through dynamical
friction, slowing down in the process.
This prediction was confirmed in $N$-body simulations (Debattista \&
Sellwood 1998, 2000; \cite[Athanassoula 2003]{Athanassoula2003};
\cite[O'Neill \& Dubinski 2003]{ONeil2003}). 
They found that bars are slowed efficiently within a few rotation
periods if a substantial density of DM is present in the region of the
bar. On the other hand, if the mass distribution is dominated by the
stellar disk throughout the inner few disk scale-lengths, then the bar
remains fast for a long time. 
These findings were challenged by \cite{Valenzuela2003} who claimed
that strong bars can rotate rapidly ($\vpd=1.3-1.7$) in
centrally-concentrated DM halos for long periods. This anomalous
behaviour was attributed by \cite{Sellwood2006} to the adopted
numerical technique which suppresses the decreasing density of
particles with angular momentum about the principal resonances usually
responsible for friction. This phenomenon could arise in nature due,
for example, to gas flows in the bar. But this is a metastable state
and mild perturbations quickly restore the frictional drag.
Thus the accurate measurement of $\om$ provides a way to discriminate
whether the central regions of the host galaxy are dominated by
baryons or by DM.

Nearly all the measured bars are found to be consistent with fast
rotators. Direct measurements of bar pattern speeds use stars as the
tracer population in SB0's and early-type spirals and gas in later
Hubble types. On the other hand, the model-dependent techniques rely
on the presence of gas and were mostly applied to late-type spirals.
Fast bars require that the disk, in which they formed, contributes
most of the rotational support in the inner parts of the galaxy. This
means that barred galaxies have maximal disks. This conclusion holds
also for bright unbarred galaxies. The comparison of the Tully-Fisher
relations for unbarred and barred galaxies shows they have a
comparable fraction of DM at a given radius (\cite[Debattista \&
Sellwood 2000]{Debattista2000}; \cite[Courteau et
al. 2003]{Courteau2003}).


\begin{thebibliography}{}

\bibitem[Aguerri et al. (2003)]{Aguerri2003} 
{Aguerri, J.~A.~L., Debattista, V.~P., \& Corsini, E.~M.}\ 
2003, \textit{MNRAS}, 338, 465 (A$+$03)

\bibitem[Aguerri et al. (2005)]{Aguerri2005} 
{Aguerri, J.~A.~L., et al.}\
2005, \textit{A\&A}, 434, 109 

\bibitem[Athanassoula (2003)]{Athanassoula2003} 
{Athanassoula, E.} 
2003, \textit{MNRAS}, 341, 1179 

\bibitem[Barazza et al. (2002)]{Barazza2002} 
{Barazza, F.\ D., Binggeli, B., \& Jerjen, H.}\ 
2002, \textit{A\&A}, 391, 823 

\bibitem[Bureau et al. (1999)]{Bureau1999} 
{Bureau, M., et al.}\
1999, \textit{AJ}, 118, 2158 

\bibitem[Corsini et al. (2003)]{Corsini2003} 
{Corsini, E.~M., Debattista, V.~P., \& Aguerri, J.~A.~L.}\ 
2003, \textit{ApJ}, 599, L29 (C$+$03)

\bibitem[Corsini et al. (2007)]{Corsini2007} 
{Corsini, E.~M., et al.}\ 
2007, \textit{ApJ}, 659, L121 (C$+$07)

\bibitem[Courteau et al. (2003)]{Courteau2003} 
{Courteau, S., et al.}\
2003, \textit{ApJ}, 594, 208 

\bibitem[Debattista et al. (2002)]{Debattista2002} 
{Debattista, V.~P., Corsini, E.~M., \& Aguerri, J.~A.~L.}\ 
2002, \textit{MNRAS}, 332, 65 (D$+$02)

\bibitem[Debattista \& Sellwood (1998)]{Debattista1998} 
{Debattista, V.~P., \& Sellwood, J.~A.}\ 
1998, \textit{ApJ}, 493, L5 
 
\bibitem[Debattista \& Sellwood (2000)]{Debattista2000} 
{Debattista, V.~P., \& Sellwood, J.~A.}\ 
2000, \textit{ApJ}, 543, 704 

\bibitem[Debattista \& Williams (2004)]{Debattista2004} 
{Debattista, V.~P., \& Williams, T.~B.}\ 
2004, \textit{ApJ}, 605, 714 (DW04)

\bibitem[Elmegreen (1996)]{Elmegreen1996} 
{Elmegreen, B.}\ 1996, in Barred Galaxies, ed. R. Buta et al., 
\textit{ASP Conf. Ser.}, (San Francisco: ASP), vol. 91,  p. 197 

\bibitem[Emsellem et al. (2006)]{Emsellem2006} 
{Emsellem, E., et al.}\
2006, \textit{MNRAS}, 365, 367 

\bibitem[Fathi et al. (2007)]{Fathi2007} 
{Fathi, K.,  et al.}\
2007, \textit{ApJ}, in press (arXiv:0708.1081) 

\bibitem[Gerssen \& Debattista (2007)]{Gerssen2007} 
{Gerssen, J., \& Debattista, V.~P.}\ 
2007, \textit{MNRAS}, 378, 189 

\bibitem[Gerssen et al. (1999)]{Gerssen1999} 
{Gerssen, J., Kuijken, K., \& Merrifield, M.~R.}\ 
1999, \textit{MNRAS}, 306, 926 (G$+$99)

\bibitem[Gerssen et al. (2003)]{Gerssen2003} 
{Gerssen, J., Kuijken, K., \& Merrifield, M.~R.}\ 
2003, \textit{MNRAS}, 345, 261 (G$+$03)

\bibitem[Hern{\'a}ndez et al. (2004)]{Hernandez2004} 
{Hern{\'a}ndez, O., et al.}\
2004, in Penetrating Bars Through Masks of Cosmic Dust, ed. D. L. 
Block et al., \textit{ASSL} (Dordrecht: Kluwer), vol. 319, p. 781

\bibitem[Hern{\'a}ndez et al. (2005)]{Hernandez2005} 
{Hern{\'a}ndez, O., et al.}\
2005, \textit{ApJ}, 632, 253 

\bibitem[Maciejewski (2006)]{Maciejewski2006} 
{Maciejewski, W.}\ 2006,  \textit{MNRAS}, 371, 451 

\bibitem[Marinova \& Jogee (2007)]{Marinova2007} 
{Marinova, I., \& Jogee, S.}\ 
2007, \textit{ApJ}, 659, 1176 

\bibitem[Merrifield \& Kuijken (1995)]{Merrifield1995} 
{Merrifield, M.~R., \& Kuijken, K.}\ 
1995, \textit{MNRAS}, 274, 933 (MK95)

\bibitem[Merrifield et al. (2006)]{Merrifield2006} 
{Merrifield, M.~R., Rand, R.~J., \& Meidt, S.~E.}\ 
2006, \textit{MNRAS}, 366, L17 

\bibitem[Noguchi (1999)]{Noguchi1999} 
{Noguchi, M.}\ 1999, \textit{ApJ}, 514, 77 

\bibitem[O'Neill \& Dubinski (2003)]{ONeill2003} 
{O'Neill, J.~K., \& Dubinski, J.} 
2003, \textit{MNRAS}, 346, 251 

\bibitem[Rand \& Wallin (2004)]{Rand2004} 
{Rand, R.~J., \& Wallin, J.~F.}\ 2004, \textit{ApJ}, 614, 142 

\bibitem[Sellwood \& Debattista (2006)]{Sellwood2006} 
{Sellwood, J.~A., \& Debattista, V.~P.}\ 
2006, \textit{ApJ}, 639, 868 

\bibitem[Tremaine \& Weinberg (1984)]{Tremaine1984}       
{Tremaine, S., \& Weinberg, M. D.}\ 1984, \textit{ApJ}, 282, L5  

\bibitem[Treuthardt et al. (2007)]{Treuthardt2007} 
{Treuthardt, P., et al.}\
2007, \textit{AJ}, 134, 1195 (T$+$07)

\bibitem[Valenzuela \& Klypin (2003)]{Valenzuela2003} 
{Valenzuela, O., \& Klypin, A.}\ 
2003, \textit{MNRAS}, 345, 406

\bibitem[Weinberg (1985)]{Weinberg1985} 
{Weinberg, M.~D.}\ 1985, \textit{MNRAS}, 213, 451 

\bibitem[Westpfahl (1998)]{Westpfahl1998} 
{Westpfahl, D.~J.}\ 1998, \textit{ApJS}, 115, 203 

\bibitem[Zimmer et al.(2004)]{Zimmer2004} 
{Zimmer, P., Rand, R.~J., \& McGraw, J.~T.}\ 
2004, \textit{ApJ}, 607, 285 

\end{thebibliography}
\end{document}